\documentclass[pra,twocolumn,superscriptaddress,showpacs,preprintnumbers, eqsecnum,amsmath,amssymb]{revtex4}
\usepackage{color}
\begin{document}
\title{Partially Unbiased Entangled Bases}
\author{A. Kalev}
\affiliation {Centre for Quantum Technologies, National University of Singapore, Singapore 117543}
\author{ F. C. Khanna}
\affiliation {Department of physics, University of Alberta, Edmonton,Alberta, Canada T6G 2J1}
\author{M. Revzen}
\affiliation {Department of Physics, Technion - Israel Institute of Technology, Haifa 32000, Israel}
%\date{\today}

\begin{abstract}
In this contribution we group the operator basis  for $d^2$ dimensional Hilbert space
in a way that enables us to relate bases of entangled states with single particle mutually
unbiased state bases (MUB), each in dimensionality $d$.
We  utilize these sets of operators to show that an arbitrary density matrix for this
$d^2$ dimensional Hilbert space system is analyzed by via $d^2+d+1$ measurements, $d^2-d$
of which involve those entangled states that we associate with MUB of the $d$-dimensional
single particle constituents. The number $d^2+d+1$ lies in the middle of the number of
measurements needed for bipartite state reconstruction with two-particle MUB ($d^2+1$) and those needed by single-particle MUB [$(d^2+1)^2$].
\end{abstract}
\maketitle
\section {Introduction}
Two orthonormal vector bases, $ {\cal B}_{1},\;{\cal B}_{2}$, are said to be mutually
unbiased bases (MUB) iff
\begin{equation}
\forall\; |u_1\rangle,\;|u_2\rangle\;\in\; {\cal B}_{1},\;{\cal B}_{2}\;\;{\rm resp.}
\;\; |\langle u_1|u_2 \rangle |={\rm constant},
\end{equation}
i.e., the absolute value of the scalar product of vectors from different bases is
constant and  independent of the vectorial labels within either basis. For a finite
dimension, $d$, Hilbert spaces the constant  is $\frac{1}{\sqrt d}$. Schwinger \cite{schwinger} was first to
emphasize that there are  more than two such bases ``that exhibit maximum degree of
incompatibility'' i.e. more than just the pair of conjugate bases such as $|x
\rangle,$  (spatial coordinates) and $ |k \rangle$ (momentum representation basis). The information theoretic oriented term ``mutually
unbiased bases''(MUB) is due to Wootters \cite{wootters1}.  For the infinite
dimensional Hilbert space case  Wootters and coworkers
\cite{wootters1,wootters2,wootters3} related the MUB's to lines in phase space. The
transcription of these notions (``lines in phase space'') to the finite, $d$,
dimensional cases may be accomplished (\cite{wootters2,wootters3,tal}) via the
eigenfunctions of the commuting operators $$X^mZ^l,\;X^{m'}Z^{l'},\;m=sm',l=sl'$$
with $m,m',l,l',s$ integers. Here $X$ and $Z$ are the Schwinger operators (SO)
\cite{schwinger}. These are discussed in the following section, and abide by the relation
$ZX=\omega XZ$, with $\omega=e^{i\frac{2\pi}{d}}$.

The maximal number of MUB was shown \cite{ivanovich} to be $d+1$. Ivanovic, \cite{ivanovich}
demonstrated that for $d=p$ (a prime) the set of $d+1$ MUB allows what is probably
the most efficient means for determining the density matrix of an arbitrary state. The
MUB analysis  attracted  great deal of research and cogent reviews are given, e.g.,
\cite{vourdas1, vourdas2,saniga}. These studies now involve abstract algebra and
projective geometry: \cite{saniga,wootters3,wootters4,klimov1,klimov2,planat1,planat2,combescure,bengtsson}.
Of particular interest to us are the articles by Planat and coworkers \cite{planat1,planat2}
who studied entangled states in conjunction with MUB sets similar to those
studied in the present paper. Similar ideas may be found also in \cite{klimov1, klimov2, klimov3, klimov4}.

In this work we aim to relate  $d$-dimensional Hilbert space MUB (with $d$ a prime)
with bases of entangled states in $d^2$-dimensional Hilbert space. To form this
relation, we classify the basis for all operators in $d^2$-dimensional Hilbert space
via SO for $d$ dimensional Hilbert space. Thus, we give a physical
interpretation for the $d^2$-dimensional Hilbert space, as two particles each in
$d$-dimensional Hilbert space. The basis for all operators in $d^2$-dimensional
Hilbert space contains $d^4$ orthogonal operators. Following the rational of
\cite{tal}, we cluster these operators into subsets of $d^2-1$ commuting operators,
and construct their common eigenvalues. We show that these states form a basis in
the $d^2$-dimensional Hilbert space. Each basis (correspond to different subset of
operators) has a simple and transparent relation to one-particle MUB. Some of the
bases constitute  maximally entangled states for the two-particle system. The relation
between these entangled states and one-particle MUB is obtained via a projection of
the one particle state onto the two particle entangled state.  Finally, we apply
our classification for the basis for all operators in $d^2$-dimensional Hilbert space for
state reconstruction (tomography). We show that one needs less measurements than are needed
for bipartite state tomography analyzed only by single-particle MUB.

In section II, utilizing formalism that stresses the (algebraic) field aspect \cite
{wootters2} applied to the approach of \cite{tal}, we obtain $d+1$ MUB for the $d$
dimensional one particle system, that form subsets each of $d-1$ {\it commuting}
orthogonal operators. These operators are selected from the $d^2$ orthogonal
complete set of operators that span the operator space. In this section cases with
dimensionality, $d$, an odd prime are discussed. Extension to $d$ powers of odd prime is
discussed in appendix A. The analysis in section II aims to setup the stage for
section III. In section III, we analyze the operator space of two-particle system,
each belong to a $d$-dimensional Hilbert space. In this section, we present the main
results of this paper. First, we associate the single-particle $d+1$ sets of operators
(each set contains $d$ orthogonal commuting operators) with two-particle bases made of
entangled states. Then we show that the $d^2$-dimensional Hilbert space of
two-particle density matrix is analyzed by $d^2+d+1$ measurements, $d(d-1)$ of
which involve  those entangled states that may be associated with single-particle
MUB. We show that particularly symmetric density matrix may be accounted for by
$d^2+1$ measurements, in close analogy with the corresponding single particle MUB
case. The numerology of the operator counting is given in Appendix B.  The last
section includes conclusions and some remarks.

\section{Mutually Unbiased Bases - Finite Dimension}

Schwinger \cite{schwinger} noted that the physics of finite dimensional, $d$, Hilbert
space is expressed by via two unitary operators, $X$, and $Z$. Thus if we label the $d$
distinct states, termed the computational basis, by
$|n\rangle,\;n=0,1,\cdots, d-1;\;|n+d\rangle=|n\rangle$, these operators are defined by:
\begin{equation}\label{def}
Z|n\rangle=\omega^{n}|n\rangle;\;X|n\rangle=|n+1\rangle,
\end{equation}
with $\omega=e^{2\pi i/d}$. They form a complete set, i.e. only a multiple
of the identity commutes with both $X,\;Z.$ We shall briefly outline a method
to utilize these operators (due mainly to \cite{tal}) to construct the $d+1$
MUB for a $d$ dimensional Hilbert space with $d$ being an odd prime. This review will be
of help in building our sets of entangled states that we shall associate with these
MUB. The computational basis vectors span the Hilbert space. All operators in this
space are expressible in terms of the $d^2$ Schwinger-operators \cite{schwinger}:
\begin{equation}\label{SO}
X^mZ^l;\;m,l=0,1,\cdots,d-1.
\end{equation}
The operators ${X^mZ^l}$ whose number is obviously $d^2$ form an orthogonal basis for
all operators in the $d$ dimensional Hilbert space,
\begin{equation}
{\rm Tr}\left[X^mZ^l\left(X^{m'}Z^{l'}\right)^{\dagger}\right]\;=\;d\delta_{m,m'}\delta_{l,l'}.
\end{equation}
This follows from Eq.(\ref{def}) which implies the commutation formula
\begin{equation}\label{com1}
XZ=\omega ZX\;.
\end{equation}
Now, let us confine ourselves to cases
wherein $m,l\;\in\ \mathbb{F}_d$ where $\mathbb{F}_d $ is a Galois field with $d$ elements.
In this case, we can relate Schwinger operators to MUB. With this aim we group Schwinger operators  (\ref{SO}) into $d+1$ sets of $d-1$ orthogonal commuting operators (which together with the identity operator form a complete operator basis for the Hilbert space). Each set of (commuting, orthogonal) operators defines a unique vector basis in Hilbert space. All the $d+1$ sets of bases form an MUB set.

Let us first consider the case $m=0$ in Eq.~(\ref{SO}). Readily, the operators  $Z^l$ with $l=0,1,\cdots,d-1$ ($l=0$ is, trivially, the identity operator) form one set of commuting and orthogonal operators. This set is diagonalized in the computational basis (c.f. Eq.~(\ref{def})). Next consider the case $m\neq 0$. In this case a unique inverse $m^{-1}$ is defined on  $\mathbb{F}_d $ and thus we can rewrite the operators (\ref{SO}) as
\begin{eqnarray}\label{power}
X^mZ^l\;=\;\omega^{\nu}(XZ^b)^{m},\\\nonumber
\end{eqnarray}
where $b=l/m,\;\nu=-\frac b2 m(m-1)$ and $b=0,1\cdots,d-1;\;\;m=1,\cdots,d-1$.
Thus, we have associated the $d(d-1)$ of the above operators in the following manner
\begin{equation}
X^mZ^{bm}\;\sim\;(XZ^b)^m.
\end{equation}
That is,  these operators differ at most by a unimodular number. Now, for a
fixed $b$, we have $(d-1)$ orthogonal and commuting operators. There are $d$ distinct such sets,
each labelled by $b$, $b=0,1,\cdots, d-1$ which are orthogonal with
\begin{equation}
{\rm Tr}\left[(XZ^b)^{m}\left((XZ^{b'})^{m'}\right)^\dagger\right]\;=d\;\delta_{b,b'}\delta_{m,m'},\;\;m,m'\ne\;0.
\end{equation}
When each $b$-labelled set of the $d-1$ orthogonal, commuting operators is supplemented
with the identity operator it constitutes  a set of $d$ unitary, orthogonal and
commuting operators. And as such, defines a vector basis for Hilbert space: The $d$ (orthonormal) vectors  diagonalize these operators in the set. (We
remark, in passing, that the basis is defined up to a choice of a phase factor which
does not affect the following results.) For each ($b$-labelled) set there exist a unique vector basis. Here, we are designating the vectors that form the basis by
$|b;c\rangle$, where the index $b$ labels the basis and $c$ that
particular vector in the basis $b$, $(b,c=0,1,\cdots, d-1)$. The expression for these states in terms of the computational basis is \cite{tal},
\begin{equation}\label{mubstate}
|b;c\rangle\;=\;\frac1{\sqrt{d}}\sum_{n=0}^{d-1}\omega^{\frac b2 n(n-1)-cn}|n\rangle,
\end{equation}
The eigenvalues are $\omega^c$.

The importance of the classification of the operators in this manner arises when considering the relation between  different sets of the vector basis.  One can readily check that these bases  form MUB, i.e.,
\begin{eqnarray}
\langle b;c|b;c'\rangle\;&=&\;\delta_{c,c'},\nonumber \\
|\langle b',c'|b,c\rangle|\;&=&\;\frac1{\sqrt{d}},\;\;b\ne b',
\end{eqnarray}
Thus, these $d$ distinct sets of bases plus the computational basis (which is mutually unbiased to all of these sets)  form the maximal number of MUB, that is $d+1$.
The proof of the last formula involves the well known \cite{schroeder} Gaussian
sums.

Since Schwinger operators, Eq.~(\ref{SO}) form an operator basis for $d$-dimensional Hilbert space, so do the set of operators $(XZ^b)^{m}$ together with $Z^b$ ($b=0,1,\cdots,d-1;\;\;m=1,2,\cdots,d-1$).
 Hence we may write an arbitrary density operator as
\begin{eqnarray}\label{rawrho}
\rho&=&\frac{1}{d}\Bigg(\sum_{m=1}^{d-1}\sum_{b=0}^{d-1}{\rm Tr}\left[\rho
(XZ^b)^m \right]\left((XZ^b)^m\right)^{\dagger} \nonumber \\
&+&\sum_{l=0}^{d-1}{\rm Tr}\left[\rho
Z^l \right]\left(Z^l\right)^{\dagger}\Bigg).
\end{eqnarray}
We are now aiming to relate this form to the expression of density operators in terms of MUB's states $|b;c\rangle$. We shall first prove the following theorem\vspace{0.2cm}\\
{\bf Completeness Theorem}\\
Let several sets of operators, each labelled by $b$, constitute of $d$
orthogonal, commuting unitary matrices, $U_{\alpha}(b)\;(\alpha=0,1,2,\cdots,d-1),$ over
a $d$ dimensional Hilbert space. We take, without loss of generality,
$U_{0}(b)=\mathbb{I}$. Let $|\nu _{b}\rangle$, $(\nu=0,1,2,\cdots,d-1)$, be a basis
that diagonalize the $d$ matrices of the $b$ set, such that
\begin{eqnarray}
\langle \nu_b|U_{\alpha}(b)|\nu'_b \rangle &=& W_{\nu}^{\alpha}(b)
\delta_{\nu,\nu'}, \\ \nonumber
|W_{\nu}^{\alpha}(b)|&=&1.
\end{eqnarray}
Then, the following (completeness) relation
\begin{eqnarray}\label{CT}
\frac {1}{d}\sum_{\alpha} W_{\nu_b}^{\alpha}(b)(W_{\nu'_b}^{\alpha}(b))^{\star}=\delta_{\nu_b,\nu'_b},
\end{eqnarray}
holds.\vspace{0.2cm}\\
{\bf Proof}\\
Lets define $d$ vectors for each set $b$,
\begin{eqnarray}
|\alpha_b \rangle&=& \frac{1}{\sqrt d}\sum_{\nu_b}W_{\nu}^{\alpha}(b)|\nu_b \rangle
\nonumber \\
\Rightarrow \langle \nu_b|\alpha_b \rangle &=&\frac{1}{\sqrt d}W_{\nu}^{\alpha}(b).
\end{eqnarray}
Now the orthogonality of $U_{\alpha}(b)$ implies
\begin{eqnarray}
d\delta_{\alpha,\alpha'}&=&{\rm Tr}\left[U_{\alpha}(b)U_{\alpha'}^{\dagger}(b)\right] \nonumber \\
&=&\sum_{\nu,\nu'}\langle \nu_b|U_{\alpha}(b)|\nu'_b \rangle \langle
\nu'_b|U_{\alpha'}^{\dagger}(b)|\nu_b \rangle
\nonumber \\
&=&\sum_{\nu}W_{\nu}^{\alpha}(b)(W_{\nu}^{\alpha'}(b))^{\star}=d\sum_{\nu} \langle \alpha'_b|\nu_b\rangle \langle\nu_b|\alpha_b
\rangle.\nonumber \\
&\therefore&\;\langle \alpha'|\alpha \rangle=\delta_{\alpha, \alpha'}.
\end{eqnarray}
Thus the sets of $d$ vectors $|\alpha_b \rangle$ form a complete orthonormal basis for
each $b$. Hence we have the completeness relation
\begin{eqnarray}
\langle \nu_b|\nu'_b \rangle&=&\delta_{\nu_b,\nu'_b}= \sum_{\alpha_b}\langle \nu_b|\alpha_b \rangle \langle
\alpha_b|\nu'_b \rangle \\\nonumber &=&\frac {1}{d}\sum_\alpha W_{\nu_b}^{\alpha}(b)(W_{\nu'_b}^{\alpha}(b))^{\star}\;
\end{eqnarray}\hfill$\Box$
 \vspace{0.2cm}

Returning to the first term of Eq. (\ref{rawrho}) - let $(XZ^b)^m$ correspond to
$U_{\alpha}(b)$, i.e. the index $m$ corresponds to $\alpha$ ($\alpha>0$) and  $U_{0}(b)=\mathbb{I}$. Since the sum in first term of Eq. (\ref{rawrho})  runs over $m>0$, for each b, we add and subtract the identity. In the second term we have that $Z^l$
goes over to $Z_{\alpha}$ - $d$ orthonormal, commuting unitary matrices with
$Z_{0}=\mathbb{I}.$ With these replacements we may write,
\begin{eqnarray}
\rho&=& \frac1d\sum_{b=0}^{d-1}\sum_{m=0}^{d-1}{\rm Tr}\left[\rho\;
U_{m}(b)\right]U_{m}^{\dagger}(b)-\mathbb{I} \nonumber \\
&+&\frac1d\sum_{m=0}^{d-1}{\rm Tr}\left[\rho Z_{m}\right]Z_{m}^{\dagger}.
\end{eqnarray}
Expressing each of these operators via its spectral representation, e.g.,
\begin{equation}
U_{m}(b)=\sum_{c=0}^{d-1}| b;c\rangle  W_{c}^{m}(b) \langle b;c|,
\end{equation}
where $| b;c\rangle$ are given in (\ref{mubstate}) and substituting into the expressions above we get, after utilizing the completeness
theorem,
\begin{eqnarray}\label{tamrho}
\rho=\sum_{b,c} |b;c
\rangle\langle b;c|\rho|b;c \rangle \langle b;c| +\sum_{n}|n \rangle\langle n|\rho|n \rangle \langle n|\;-\mathbb{I}\nonumber\\
\end{eqnarray}
This is an expression for $\rho$  in terms of probabilities. The number
$\langle b;c|\rho|b;c \rangle$ corresponds to the probability to find the state $|b;c \rangle$ in $\rho$. We see that, as shown in
\cite{ivanovich}, the state of the system, i.e. $\rho$, is obtainable via the $d+1$
measurements, where the
measurement of a unitary operator is understood by the measurement of the two
commuting hermitian operators,
\begin{eqnarray}\label{meas}
\hat{M}_1&=&\hat{X}\hat{Z}^b+(\hat{X}\hat{Z}^b)^{\dagger} \nonumber \\
\hat{M}_2&=&i (  \hat{X}\hat{Z}^b-(\hat{X}\hat{Z}^b)^{\dagger} ).
\end{eqnarray}
Each of these measurements
yields the $d-1$ independent probabilities  (since the sum of the
probabilities adds to one). This gives $(d+1)(d-1)=d^2-1$ numbers that determine the
density matrix. It should be noted that this holds since the operators are non-degenerate.

\section{Entanglement associated with MUB}

We now wish to consider two $d$ dimensional particle state. The Hilbert space
dimensionality is now $d^2$ while the operator space is $d^4$ dimensional. Our aim
is to utilize the single particle MUB given above to construct two particle entangled
state bases that, thereby, may be viewed as associated with the MUB.

We first wish to state what we consider as a self evident sufficient
condition for entanglement: Given  single particle (particle $\mu$) operators $A$ and
$A'$ which do not commute: $AA'=\alpha A'A$, and similarly, given the non-commuting
operators $B$ and $B'$ for a second particle (particle $\nu$): $BB'=\beta B'B$. Let
$\alpha,\;\beta$ be scalars with $\alpha \beta=1$. The two two-particle operators
$AB$ and $A'B'$ commute. We now assert that the common eigenfunction of these is an
entangled state - we are unaware of a formal proof of this which we consider as
obvious and {\it use it as a guide} for constructing entangled states.

Returning to our two-particle Hilbert space. The discussion of the previous section
implies that the $d^4$ unitary orthogonal operators,
\begin{equation}\label{general}
(X^{m_1}Z^{l_1})_\mu (X^{m_2}Z^{l_2})_\nu,\;\;m_1,l_1,m_2,l_2=0,1,\cdots,d-1,
\end{equation}
form a complete orthonormal operator basis. Restricting our consideration to
$m_1,l_1,m_2,l_2\;\in\ \mathbb{F}_d,$ allow us to re-construct and classify the $d^2$ operators in (\ref{general}) as follows
\begin{eqnarray} \label{sets}
(XZ^{b_1})_{\mu}^{m_1}(XZ^{b_2})_{\nu}^{m_2},&& \;(d(d-1))^2\;operators \nonumber \\
Z^{b_1}_{\mu}(XZ^{b_2})_{\nu}^{m_2}&& \;d\cdot d(d-1)\;operators \nonumber \\
(XZ^{b_1})_{\mu}^{m_1}Z^{b_2}_{\nu}&&\;d\cdot d(d-1)\;operators \nonumber \\
Z_{\mu}^{b_1}Z_{\nu}^{b_2}&&\;d^2\;operators\;,
\end{eqnarray}
here $b_1,b_2=0,1,\cdots,d-1$ and $m_1,m_2=1,\cdots,d-1$.
This construction is carried out with an exact analogy to the one-particle case (c.f. Eq.~(\ref{power})). One readily verifies that the above $d^4$ operators are orthogonal and thus span the operator space.

First, let us note that these operators are diagonalized in the one-particle MUB basis that was introduced in section II. Alas, these bases do not form an MUB set on the two-particle Hilbert space. Thus, for example, the basis which diagonalize $Z^{b_1}_{\mu}(XZ^{b_2})_{\nu}^{m_2}$ is $\{|n\rangle_\mu|b_1;c\rangle_\nu\}$, while the operators  $Z^{b_1}_\mu Z^{b_2}_\nu$ are diagonalize in $\{|n'\rangle_\mu|n\rangle_\nu\}$. These bases are clearly not MUB. However, in this construction we do find sets of MUB for the two-particle Hilbert space that is directly constructed from the one-particle MUB. The basis which diagonalize $Z^{b_1}_{\mu}(XZ^{b_2})_{\nu}^{m_2}$ (i.e., $\{|n\rangle_\mu|b_1;c\rangle_\nu\}$) and the basis that operators  $(XZ^{b_1})_{\mu}^{m_1}Z^{b_2}_{\nu}$ -- $\{|b_1;c\rangle_\mu|n\rangle_\nu\}$ -- are MUB. Another set of MUB are the basis that diagonalize  $(XZ^{b_1})_{\mu}^{m_1}(XZ^{b_2})_{\nu}^{m_2}$ ($\{|b_1;c\rangle_\mu|b_2;c'\rangle_\nu\}$) and the basis that diagonalize $Z_{\mu}^{b_1}Z_{\nu}^{b_2}$.

We  show that some of the above operators are diagonalized by entangled states that form two-particle MUB and have a close relation to a single-particle MUB.

To show this, we rewrite the first set in Eq.~(\ref{sets}) as
\begin{eqnarray}\label{seta}
&&\Big(X_\mu(X^{s}Z^{sb})_{\nu}(Z^s_{\mu}Z^{-1}_{\nu})^m\Big)^{m_1},\;  \nonumber \\
&& s,m_1=1,2,\cdots,d-1,\; b,m=0,1,2,\cdots,d-1,
\end{eqnarray}
where the parameters $s$,$m$ and $b$ are defined as $sm_1\equiv m_2\;,sm\equiv b_1$ and $b_2\equiv bm_2-mm_1$. This set, for fixed $s$ and $b$, contains $d(d-1)$ orthogonal, commuting
operators.(Allowing $s$ and $b$ to vary, the sets contain $[d(d-1)]^2$ orthogonal
operators.) Now consider $d$ orthogonal operators extracted from the last set of
Eq.(\ref{sets}) in the following way: We partition the set into two sets by,
($r,s=1,2,\cdots,d-1$)
\begin{eqnarray}\label{setb}
Z^{sr}_{\mu}Z^{-r}_{\nu}&&\;(d-1)^2\;operators \nonumber \\
Z^s_{\mu}\;&&\;(d-1)\;operators,\nonumber \\Z^s_{\nu}\;&&\;(d-1)\;operators,
\end{eqnarray}
these, with the identity operator, add up to the set of $d^2$ orthogonal, commuting
unitary operators. For fixed $s$, the $(d-1)$ operators $Z^{sr}_{\mu}Z^{-r}_{\nu}$
commute with the $d(d-1)$ operators given in Eq.(\ref{seta}) defined above to form,
 with the identity, $d^2$ orthogonal commuting operators in our $d^2$
dimensional Hilbert. Thus these define the representations wherein they are
diagonal. In terms of the computational basis for the two particles, these stats are
given by
\begin{equation}\label{entmub}
|b,s;c_1,c_2 \rangle_{\mu,\nu}=\frac{1}{\sqrt d} \sum_{n}
\omega^{(\frac{s^2b}{2})n(n-1)-c_1n}|n \rangle_{\mu}|sn+c_2 \rangle_{\nu}.
\end{equation}
Then it can be checked that ($c_1,c_2=0,1,2\cdots,d-1$, we shall eschew the label $\mu\nu$ below)
\begin{eqnarray}
Z^{s}_{\mu}Z^{-1}_{\nu}|b,s;c_1,c_2\rangle&=&\omega^{-c_2}|b,s;c_1,c_2\rangle\rangle \nonumber\\
(X)_{\mu}(X^{s}Z^{sb})_{\nu}|b,s;c_1,c_2\rangle&=&\omega^c|b,s;c_1,c_2 \rangle,
\end{eqnarray}
\noindent $c=sbc_2$. These states are entangled states (in accord with our guide),
they are orthonormal and each set spans the $d^2$ dimensional Hilbert space. Since
each operator set is labelled with the two indices, $s$ and $b$, we have $d(d-1)$ distinct
sets which constitutes a $d^2$ dimensional base composed of entangled states.
Within these sets, $d-1$ of them are MUB. By direct calculation one
obtains,
\begin{eqnarray}
|\langle b,s;c_1,c_2|b',s';c_1',c_2'\rangle|&=&\delta_{c_1,c_2}\;\;b=b',\;s=s',\nonumber \\
&=&\frac{\delta_{c_2,c_2'}}{{\sqrt d}}\;\;s=s',\;b \ne b',\;\nonumber \\
&=&\frac{1}{d}\;\;s \ne s'\;b \ne b'.
\end{eqnarray}
Thus, bases with distinct $s$ and $b$ are MUB (i.e. d-1 entangled state bases) while
common $s$ and $b$ defines a complete orthonormal $d^2$ dimensional basis.

Next, we wish to relate these entangled state to a one-particle MUB.
To convey the idea we focus on the set defined by $s=1$, and without loss of generality we consider $c_2=0$:
\begin{equation}\label{nn}
|b,s=1;c_1,c_2=0\rangle_{\mu,\nu}=\frac{1}{\sqrt{d}}\sum_{n}\omega^{\frac b2 n(n-1)-c_1n}|n\rangle_\mu |n\rangle_\nu\;.
\end{equation}
Suppose that the $\mu$ particle is found in one of the  states which belong to the MUB that were introduced in section II. Then, the reduced state for the $\nu$ particle is (up to normalization)
\begin{equation}\label{nnp}
\,_\mu\langle b_1;c|b,s=1;c_1,c_2=0\rangle_{\mu\nu}=\frac{1}{\sqrt d}|b-b_1;c_1-c\rangle_\nu\;.
\end{equation}
This is the exact meaning of associating entangled states with single particle MUB.
For the case $b_1>b$, the basis label for particle 2, $b_2=b-b_1$, is negative and should be defined as modulus $d$.
Note that for an odd dimension $d$, $b_1$ is always different from $b_2$. This ensures that, after measurement, states of the two particles belong to different MUB.
The projection given in Eq. (\ref{nnp}) corresponds to  Hermitian operators which were introduced in Eq. (\ref{meas}).

Finally, we gather the operators that provide a basis for the $d^2$-dimensional Hilbert space to state reconstruction. For this, we may use  the $d^4$ unitary operators given in
Eq.(\ref{general}) as is rewritten via Eqs.~(\ref{sets}), (\ref{seta}), and( \ref{setb})  to express an arbitrary two-particle density matrix as
\begin{eqnarray} \label{rho2p}
 \rho &=&\sum_{b,s}\sum_{m,n}{\rm Tr}\Bigg[\rho
\Bigg(X_{\mu}(X^sZ^{sb})_{\nu}(Z^s_{\mu}Z^{-1}_\nu)^m\Bigg)^{n}\Bigg]
\nonumber\\&\times&\bigg[X_{\mu}(X^sZ^{sb})_{\nu}(Z^s_{\mu}Z^{-1}_\nu)^m\bigg]^{n\dagger} \nonumber \\
&+&\sum_{s,n}\bigg(tr\rho\bigg[(Z^s)_{\mu}(Z^{-1})_{\nu}\bigg]^n\bigg)\bigg[(Z^s)_{\mu}
(Z^{-1})_{\nu}\bigg]^{n\dagger} \nonumber \\
&+&\Bigg[\sum_{n,m}\sum_{s}\bigg(tr\rho
Z_{\mu}^n(XZ^m)_{\nu}^s\bigg)\bigg(Z_{\mu}^n(XZ^m)_{\nu}^s \bigg)^{\dagger}
\\
&&+\sum_{n}\bigg( tr\rho Z_{\mu}^{n} \bigg)Z_{\mu}^{n\dagger}\;+\;
\mu\;\Longleftrightarrow\;\nu \Bigg]\;-\;\mathbb{I}.
\end{eqnarray}
Using what we termed the ``completeness'' theorem that is proven in section II we
rewrite this as (we eschew the particle number label viz. $\mu,\nu$ as being obvious in
the expression below), where
\begin{eqnarray} \label{rhoexplicit}
\rho &=& -\mathbb{I}+\sum_{b,s}\sum_{c_1,c_2}\langle b,s;c_1,c_2|\rho|b,s;c_1,c_2
\rangle \\\nonumber
&&\;\;\;\;\;\;\;\;\;\;\;\;\;\;\times\;\; |b,s;c_1,c_2\rangle_{\mu,\nu}\langle b,s;c_1,c_2| \\\nonumber
&-&(d-1)\sum_{n,n'} \langle
n,n'|\rho|n,n'\rangle \times |n\rangle_{\mu}\langle n|\otimes|n'\rangle_{\nu}\langle n'|  \\\nonumber
&+&\sum_{b=0}\sum_{n,c}\langle
b;c|\langle n|\rho|n\rangle |b;c\rangle\times |n\rangle_{\mu}\langle n|\otimes |b;c\rangle_\nu  \langle b;c|  \\\nonumber
&+&\sum_{b=0}\sum_{n,c}\langle n
|\langle b;c|\rho|b;c\rangle |n\rangle\times |b;c\rangle_{\mu}\langle b;c|\otimes |n\rangle_\nu  \langle n| \;,
\end{eqnarray}
where $s=1,2,\cdots,d-1,$ $b,n,n',c_1,c_2=0,1,\cdots,d-1$.
In the above expression, $\rho$ is given in terms of probabilities. Thus Eq.~(\ref{rho2p}) presents an operational scheme for two-particle state reconstruction  which is based on Schwinger operators of one-particle. For example, the probability of finding the system whose state is $\rho$ in the (entangled) state  $|b,s;c_1,c_2\rangle$ two
readings are required: that pertaining to measuring
$X_{\mu}(X^sZ^{sb})_{\nu}$ and one that gives the value of measuring
$Z^s_{\mu}Z^{-1}_{\nu}$. We consider these two readings as one measurement as it
refers to the same sample. With this
in mind an arbitrary density matrix is given in terms of $d^2+d+1$ measurements: The
first term implies $d(d-1)$ measurements, the second one and the third and fourth $d$
measurements each. The total number of measurements exceeds the optimal number, i.e., $d^2+1$, since not all the bases in this construction are MUB. However, the number of measurements in this scheme is less than one would have needed by considering only single-particles MUB, that is, $(d+1)^2=d^2+2d+1$ measurements.

\section{Conclusions and Remarks}

Mutual unbiased bases for one particle states in $d$ dimensions may be
characterized by states that diagonalize a set of $d$ commuting orthogonal unitary
operators which, in turn, is one of $d+1$ such sets. The central attribute of these is
that a measurement ascertaining the particle to be in a state $|u\rangle$ of one
such basis implies that it to be with equal likelihood in any of the states
$|v\rangle$ of any of the other bases. Maximally entangled state of  two particles
central attribute is that partial tracing of the coordinates of one of the particles
(i.e. performing non selective measurements on it) leaves the other particle with
equal likelihood in any state. We relate these two notions by constructing $d(d-1)$ entangled states base vectors that form an MUB as well as relate to the one-particle MUB. We also gave an operational content for this interpretation. We showed that an arbitrary two particle density matrix accounting for Hilbert space dimensionality $d^2$, is accountable via
$d^2+d+1$ measurements where $d(d-1)$ of the measurements involve entangled states  which we related to the single particle MUB.

\begin{acknowledgments}
FCK acknowledges the  support of NSERC and the hospitality of the Physics department, Technion - Israel Institute of Technology where part of this work was carried out. Informative discussions with our colleagues Professors J. Zak and A. Mann are gratefully acknowledged.
\end{acknowledgments}

\section*{Appendix A: Operator count at finite dimensionality}

The numerology involves two counts. The first shows that our groupings of the
operators retains all of them and each just once. The second involves the algebraic
addition and subtractions of operators to obtain groupings each of which involves
$d^2-1$ (not including the identity) orthogonal and {\it commuting} operators
eligible for the application of the completion theorem, section II. We begin with
the first demonstration: Eq. (\ref{seta}) contains $d^2(d-1)^2$ orthonormal
operators. Next we add the two sets of Eq.(\ref{setb}) that contain $d^2-1$
orthonormal operators. Now we add the two sets that constitute the third and fourth
lines of Eq. (\ref{sets}) these these contain $2d^2(d-1)$ orthonormal operators. The
sum total of these with
the identity added is $d^4.$ QED.\\
The groupings of sets of $d^2-1$ of commuting (and orthogonal) sets amendable for
the applicability of the completion theorem is done as follows: first we add to each
of the $d(d-1)$ commuting operators labelled by distinct $b$ and $s$ the set of $d-1$
operators $(Z^s_{\mu} Z^{-1}_{\nu})^n,\;n=1,2,\cdots,d-1$. The combined set is composed
of $d^2-1$ orthogonal and commuting operators (to which, when supplemented with the
identity, we associate the $d^2$ dimensional orthonormal basis
$|b,s;c_1,c_2\rangle).$ To compensate we subtract this added set and then combine
this subtracted set with the rest of the operators. Now we add to it
$-(d-1)\sum_{n=1}\left(Z^n_{\mu}+Z^n_{\nu}\right),$ these together (supplemented
with the identity) form $d-1$ sets each with $d^2$ commuting and orthogonal unitary
operators. Correcting for this subtraction we add to each of d sets ($b=0,1,\cdots,d-1$)
of the form $\sum_{n=0,m=1}Z^n_{\mu}(XZ^b)^m_{\nu}$ to get again sets each of which
is made up (with the identity) of $d^2$ commuting and orthogonal unitary operators.
%This may be summarized by ($b=0,1,\cdots,d-1$),
%\begin{eqnarray}
%&&\sum_{m,n,n'm'}X_{\mu}^nZ_{\mu}^mX_{\nu}^{n'}Z_{\nu}^{m'},\;\;m,n,n',m'=0,1,\cdots,d-1.
%\nonumber \\
%&&\sum_{b,s}\left[\sum_{b_1,n}\left(XZ^{b_1}_{\mu}X^sZ^{sb_2}_{\nu}\right)^n
%\end{eqnarray}
%
%\begin{eqnarray}
%&&\sum_n \left(X^s_{\mu}Z^{-1}_{\nu}\right)^n+\mathbb{1}\right]\;-
%\;(d-1)\left[\sum_{m,n}Z^n_{\mu}Z^m_{\nu}\right]+ \nonumber \\
%&&\sum_{b} \left[\sum_{n=0,m=1}Z^n_{\mu}\left(XZ^b)^m\;-\;\sum_{n=0}Z^n_{\mu}\right]+
%\nonumber \\
%&&\sum_{b}
%\left[\sum_{n=0,m=1}Z^n_{\mu}\left(XZ^b)^m\;-\;\sum_{n=0}Z^n_{\mu}\right]-d^2\mathbb{1}.
%\end{eqnarray}
%Here each square bracket set involves $d^2$ commuting orthogonal unitary
%operators.

\section*{Appendix B: Entanglement of MUB states at finite dimensionality}

Our method for the finite dimensional may be applied  to the case where the states'
label are finite field variables. This is the case for dimensionality $d$, with $d=p^n$, $p$
an odd prime \cite{klimov1,klimov2,planat1,planat2}. The MUB are given by, using our
notation, ($\omega_p=e^{i\frac{2\pi}{p}}$):
\begin{equation} \label{mub}
 |b;c\rangle\;=\;\frac{1}{\sqrt{d}}\sum_{n\in\mathbb{F}_d}{\omega_{p}^{tr[\frac b2 n^2
 +cn]}}|n\rangle.
\end{equation}
Here $b,c,n\; \in\; \mathbb{F}_d,$  $\mathbb{F}_d$ is Galois field with $d$ elements, $|n\rangle$ vector in the computational basis (labelled with an element of
the field); and $tr[\alpha]=\alpha+\alpha^p+\alpha^{p^2}+\cdots+\alpha^{p^{n-1}}$. The
trace ,tr, is a mapping with $\alpha
\;\in\;\mathbb{F}_d;\;\;tr{\alpha}\;\epsilon\;\mathbb{F}_p$. Now a basic
property of trace is: $tr[\alpha+\beta]=tr[\alpha]+tr[\beta].$ (The factor $\frac12$ in the exponent means the solution of $2x=1\;\in\;\mathbb{F}_d.$) Using our
entanglement scheme (cf. \cite{planat1,planat2}), i.e. entangling within the
computational basis, the entangled state corresponding to the generic MUB state Eq.
(\ref{mub}) is,
\begin{eqnarray}
|b;c\rangle_{\mu\nu}&=&\frac{1}{\sqrt d}\sum_{n\epsilon\mathbb{F}_d}{\omega_p^{tr[\frac{b}{2}n^2+cn]}} |n\rangle_\mu
|n\rangle_\nu,  \\\nonumber
 &=&\frac{1}{\sqrt d}\sum_{n\epsilon\mathbb{F}_d}{\omega_p^{tr[\frac{b_1}{2}n^2+c_1n]}
\omega_p^{tr[\frac{b_2}{2}n^2+c_2n]}}|n\rangle_\mu |n\rangle_\nu,
\end{eqnarray}
with $b_1+b_2=\;b,\;\;\;c_1+c_2=c$. Thus projecting the first particle to (any of
the) $b_1,\; c_1$ labelled state results in projecting the second particle to
$|b_2,c_2 \rangle_\nu$ state.
\begin{equation}
\;_\mu\langle b_1;c_1|b;c\rangle_{\mu\nu}\;=\;\frac{1}{\sqrt d}|b_2;c_2\rangle_\nu.
\end{equation}

\end{document}